\documentclass[superscriptaddress,aps,prX,twocolumn,showpacs,nofootinbib,longbibliography,notitlepage,floatfix]{revtex4-2}

\usepackage[latin9]{inputenc}
\usepackage{mathtools}
\usepackage{amsbsy}
\usepackage{amstext}
\usepackage{amsmath}
\usepackage{amssymb}
\usepackage{txfonts}
\usepackage{color}
\usepackage{graphicx}
\usepackage{wrapfig}
\usepackage{float}
\usepackage[normalem]{ulem}
\usepackage{physics}
\usepackage{bbold}
\usepackage{cancel}

\usepackage[unicode=true,pdfusetitle,
 bookmarks=false,
 breaklinks=false,
 pdfborder={0 0 1},
 backref=false,
 colorlinks=false]
 {hyperref}

\hypersetup{
 bookmarksnumbered=false,
 bookmarksopen=false}

\makeatletter

\@ifundefined{textcolor}{}{
 \definecolor{BLACK}{gray}{0}
 \definecolor{WHITE}{gray}{1}
 \definecolor{RED}{rgb}{1,0,0}
 \definecolor{GREEN}{rgb}{0,0.7,0}
 \definecolor{BLUE}{rgb}{0,0,1}
 \definecolor{CYAN}{cmyk}{1,0,0,0}
 \definecolor{MAGENTA}{cmyk}{0,1,0,0}
 \definecolor{YELLOW}{cmyk}{0,0,1,0}
}

\newcommand{\beq}{\begin{equation}}
\newcommand{\eeq}{\end{equation}}
\newcommand{\beqa}{\begin{eqnarray}}
\newcommand{\eeqa}{\end{eqnarray}}

\makeatother

\begin{document}	

\title{Microscale Sensing with Strongly Interacting NV Ensembles at High Fields}

\author{Ainitze Biteri-Uribarren}
\email{Contact author: ainitzebiteriuribarren@gmail.com}
\affiliation{Department of Physical Chemistry, University of the Basque Country UPV/EHU, Apartado 644, 48080 Bilbao, Spain}
\affiliation{EHU Quantum Center, University of the Basque Country UPV/EHU, Bilbao, Spain}

\author{Ana Martin}
\affiliation{Department of Physical Chemistry, University of the Basque Country UPV/EHU, Apartado 644, 48080 Bilbao, Spain}
\affiliation{EHU Quantum Center, University of the Basque Country UPV/EHU, Bilbao, Spain}

\author{Jorge Casanova}
\affiliation{Department of Physical Chemistry, University of the Basque Country UPV/EHU, Apartado 644, 48080 Bilbao, Spain}
\affiliation{EHU Quantum Center, University of the Basque Country UPV/EHU, Bilbao, Spain}

\begin{abstract}
Advances in sensing devices that utilize nitrogen-vacancy (NV) center ensembles in diamond are driving progress in microscale nuclear magnetic resonance spectroscopy. Utilizing quantum sensing techniques in the high-field regime significantly boosts sensitivity by increasing thermal polarization and improves spectral quality via enhanced energy shifts. Compatible with the latter, a straightforward manner to further raise sensor sensitivity is to increase NV concentration, although this intensifies detrimental dipole-dipole interactions among NVs. In this Letter, we present a method for detecting NMR signals in high-field scenarios while effectively suppressing dipole-dipole couplings in the NV ensemble. Thus, this approach enhances sensitivity by combining highly doped diamond substrates and elevated magnetic fields.
\end{abstract}

\maketitle

\section{Introduction} Quantum sensors have emerged as tools for detecting magnetic signals in different regimes, spanning pure and applied science~\cite{Degen17}. Notably, in the realm of nuclear magnetic resonance (NMR) spectroscopy~\cite{Abragam61,Levitt08}, conventional inductive detection techniques struggle with inherent low sensitivity, limiting their application to millimetre sized samples~\cite{Allert22}. On the contrary, quantum sensor-based NMR shows great potential for applications involving minimal sample sizes~\cite{GlennBucher2018, Schmitt2017, Allert22, Degen2014, Briegel2024} such as high-throughput chemistry~\cite{Whitesides2006, Mennen2019}, single-cell biology~\cite{Neuling2023}, and two-dimensional material science~\cite{Gibertini2019}. In this context, nitrogen-vacancy (NV) color defects in diamond~\cite{Doherty13, Degen2014, Allert22} have gained popularity owing to their performance at ambient conditions, making microscale ensembles of NVs a viable solution to investigate picoliter samples via NMR techniques.

Recently, efforts have concentrated on synthesizing diamonds with a large NV concentration to further enhance NV-based NMR sensitivity~\cite{Chakraborty2019, Hughes2024, Healey2021, Dhungel2024, Choi2020, droid, Zhou2023, Tyler2023, Arunkumar2023, Zhou2024}. Having more NV centers lowers the system's noise floor, resulting in a higher signal-to-noise ratio (SNR)~\cite{Taylor2008, Barry2020}. In light of this, researchers  are designing various mechanisms based on dynamical decoupling techniques (DD) to mitigate detrimental dipole-dipole interactions that arise in highly NV-doped diamonds~\cite{Choi2020, Zhou2023, Tyler2023, Arunkumar2023, Oon2024}. However, these techniques do not extend to the high-field regime, which is a particularly relevant domain for various applications. Among other benefits, elevated fields amplify the sample-emitted signal due to increased nuclear thermal polarization and also improve the spectral resolution of chemical shifts~\cite{AERIS, DRACAERIS, J_INSECT, Solid_state, Meinel2023}.

In this work, we introduce SHIELD (Sensing at High field Improved via ELusion of Dipole-dipole), a microwave DD sequence specifically designed for densely packed NV ensembles, capable of operating at high-field conditions. Our sequence simultaneously suppresses dipolar interactions among NVs while coupling the sensor to the high-field NMR signal.  Besides, it is robust against common experimental errors such as strain in the diamond lattice, couplings with other impurities in the host substrate, and errors on the applied control drivings. Additionally, our sequence can be incorporated to quantum heterodyne protocols that enable high-resolution sensing~\cite{Schmitt2017, GlennBucher2018}.
\begin{figure}[h]
\includegraphics[width= 0.75 \linewidth]{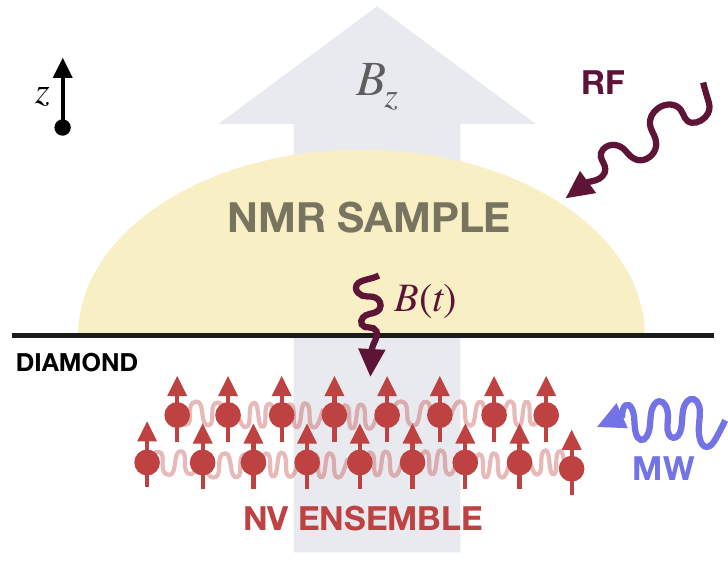}
\caption{Illustration of the setup: A [111] cut diamond hosts a dense ensemble of NV centers, exhibiting strong dipole-dipole interactions. A target sample containing nuclear spins is placed on the diamond surface, with a strong magnetic field in the order of Teslas is applied along the $z$-axis. Nuclear spins rotate at moderate speed due to an RF field resulting into a slowly oscillating signal, $B(t)$, that encodes sample information. Simultaneously, NV centers are controlled using an MW driving.}\label{fig:setup}
\end{figure}

\section{METHODS}\subsection{Theoretical framework} We consider a nuclear spin sample placed on top of a diamond with embedded NV centers, in the presence of a large external magnetic field, $B_z$; see Fig.~\ref{fig:setup}. The target NMR signal $B(t)$ is generated by driving the sample with radio frequency (RF). In this manner, even in the high-field regime, the target signal oscillates at a slow rate, in particular, at the nuclear Rabi frequency ($\sim 100$ kHz)~\cite{AERIS, DRACAERIS, J_INSECT, Solid_state} . Our goal then, is the detection of this slowly oscillating  $B(t)$ with a highly dense NV ensemble. 

The Hamiltonian describing the system is

\begin{align} 
  \frac{H}{\hbar}=&\sum_{i=1}^N \left[ D (S^z_i)^2 + (|\gamma_e| B_z+\xi_i) S^z_i \right] + \nonumber \\
   			&\sum_{i=1}^N |\gamma_e| B(t) S^z_i+ \sum_{i<j}^N H_{dd}^{ij}+\sum_{i=1}^N \sqrt{2} \Omega \cos{(\omega_c t)} S_i^{\phi}, 
\label{eq:Hamiltonian}
\end{align}

where $S_i^{\phi}=\cos{\phi} S_i^x + \sin{\phi} S_i^y$, while $S_i^{x,y,z}$ are the \mbox{spin-1} operators. The first term in Eq. \eqref{eq:Hamiltonian} accounts for the free energy of $N$ NV centers, determined by the zero-field splitting $D/(2\pi) = 2.88$ GHz, the electron gyromagnetic ratio $|\gamma_e|/(2\pi) = 28.024$ GHz/T, the external magnetic field $B_z$, and the site dependent energy shifts $\xi_i$, which can introduce variations up to MHz~\cite{Lukin} in the resonance energy of the distinct NV centers in the ensemble. The second term describes the interaction between the NV ensemble and the NMR signal $B(t)$, which we define as $B(t)=B_t \cos(\omega t+\phi_t)$; and the third term, $\sum_{i<j}^N H_{dd}^{ij}$, corresponds to the dipole-dipole interaction among NVs, see Supplemental Material (SM)~\cite{SM} for an explicit expression of $H_{dd}^{ij}$. The final term corresponds to the microwave (MW) control driving with amplitude $\Omega$ and frequency $\omega_c$.

The Hamiltonian of the system (Eq.~\ref{eq:Hamiltonian}), in a rotating frame with respect to $H_0=\sum_{i=1}^N \omega_c S_i^z$ reads 
\begin{equation}\label{eq:IP}
\frac{H_I}{\hbar}=\sum_{i=1}^N \xi'_i s^z_i + \sum_{i=1}^N \left[ \Delta s^z_i + \Omega s_i^{\phi} \right]+ \sum_{i=1}^N \gamma_e B(t) s^z_i+ \sum_{i<j}^N \bar{H}_{dd}^{ij},
\end{equation}
where $\Delta= (D-|\gamma_e| B_z )-\omega_C$ is the detuning on the MW control; and  $\bar{H}_{dd}^{ij} = d_{ij} [ s_i^z s_i^z - (s_i^x s_j^x +s_i^y s_j^y)]$, while $\xi'_i= \xi_i+ \sum_{i\neq j}d_{ij}/2$. Note that spin operators $s_i^{x,y,z}$ are now spin-1/2 operators; for simplicity, the Hamiltonian is restricted to the subspace spanned by $\{\ket{b_1}\otimes\cdots\otimes\ket{b_N}\mid b_j\in\{0,-1\}\}$, neglecting spin occupations in $\ket{1}_j$ (see SM~\cite{SM} for details). 

We start by analyzing the role of dipole-dipole interactions among NV detectors in the widely used Carr-Purcell-Meiboom-Gill (CPMG)  protocol~\cite{Meiboom1958}. As in NV-based NMR experiments, the expectation value $s^z$ of NV centers is inferred, we compute the latter via Eq.~\eqref{eq:IP}. In particular, for the CPMG protocol, the system dynamics is described with Eq.~\eqref{eq:IP} by setting $\Delta=0$ while $\Omega$ is deployed stroboscopically. Then, after applying CPMG over a cluster of $q$ interacting NVs, we have (see SM~\cite{SM})
\begin{eqnarray}\label{eq:spectrum}
\langle {\textstyle \sum_{i=1}^q} s_i^z \rangle  \approx \frac{2\gamma_e B_t t_s}{\pi }\cos(\phi_t) \  \langle \sum_{i=1}^q s_i^x \rangle_{\rho^\ast},
\end{eqnarray}
where $t_s$ is the CPMG  sequence duration, and $\rho^\ast$  the NV cluster state evolved only under dipole-dipole interactions. Hence, the term  $\langle \sum_{i=1}^q s_i^x \rangle_{\rho^\ast}$  depends on the dipole-dipole interactions, and it can be bounded as  $\big|\langle {\textstyle \sum_{i=1}^q} s_i^x \rangle_{\rho^\ast}\big|\leq q/2$, achieving its maximum $q/2$ in the absence of dipole-dipole couplings. To illustrate this concept, we analytically solved the case involving two NV centers (see SM~\cite{SM}).

In a scenario with an NV ensemble composed of $Q$ clusters, each containing $q_k$ NV centers (where $k=1$, ..., $Q$), the total NV signal is the sum from all clusters: $\sum_{k=1}^Q \langle \sum_{i=1}^{q_k} s_i^z \rangle$. Then, the contribution of each cluster to the total NV signal has a random factor in the range $[-q_k/2, q_k/2]$, depending on the structure of dipole-dipole terms within the cluster. This means that ($i$) Each cluster, most likely, generates a weaker signal compared to that without dipole-dipole interactions, and ($ii$) In the total NV ensemble signal different clusters can interfere destructively, making signal detection challenging. This underlines the need for protocols that suppress dipolar interactions to boost NV-based quantum sensing.

\begin{figure*}
\includegraphics[width= 0.9 \linewidth]{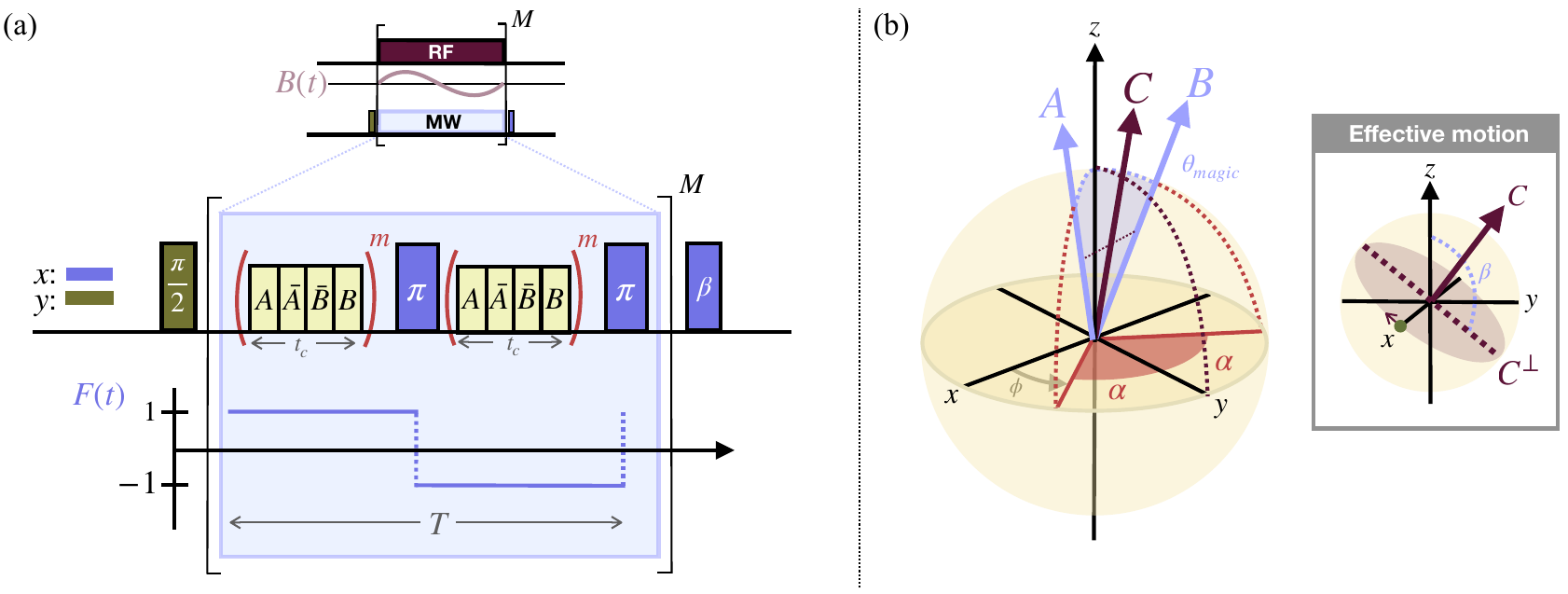}
\caption{Scheme of the protocol. (a) At the top, continuous RF driving induces Rabi oscillations of the sample nuclei, which results in the NMR signal $B(t)$. At the bottom, the blue square represents the MW pattern of our sequence SHIELD, which simultaneously captures $B(t)$ while suppressing dipole-dipole interactions. SHIELD includes an initial and final pulse, and in between, $M$ repetitions of two $m$ blocks of rotations around the axes $A$, $\bar{A}$, $\bar{B}$, and $B$, interspersed by $\pi$-pulses along the x-axis. These pulses imprint the function $F(t)$ with periodicity $T$ on the effective operators $\sigma^C_j$. (b) Relevant direction and angles. In the XY plane, the axes $A$ and $B$ are rotated by angles $-\alpha$ and $\alpha$ from the $y$-axis about the $z$-axis, and both axes --$A$ and $B$-- form the magic angle with the $z$-axis. The bisector of $A$ and $B$ defines direction $C$. The grey box inset illustrates the effective dynamics of the NV spins: the initial pulse polarizes the NV centers along the $x$-axis (green dot), then the MW scheme slightly rotates the initial state around $C$, inducing changes in $\langle S^{C\perp} \rangle$, where $C^{\perp} =C \times x$. For readout, the accumulated spin polarization along $C^{\perp}$ is transferred to $z$-axis via the final pulse of angle $\beta$ along the $x$-direction.}\label{fig:scheme}
\end{figure*}

\subsection{The protocol} Our protocol --SHIELD--, integrates MW driving blocks that continuously mitigate the effect of NV-NV interactions. These blocks are interspaced by $\pi$-pulses, which suppress disorder terms $\xi'_i$ whilst simultaneously coupling the NV ensemble to $B(t)$; refer to Figure~\ref{fig:scheme}~(a) for a scheme of our sequence. 

The MW driving blocks generate concatenated \(2\pi\) rotations of the NV centers around the axes \(A\), \(B\), \(\bar{A} = -A\), and/or \(\bar{B} = -B\), realized by fixing the appropriate phase \(\phi\) and setting the MW detuning to \(\Delta = \pm\Omega / \sqrt{2}\); see Eq.~\eqref{eq:IP}.  This specific choice of detuning continuously symmetrizes the dipole-dipole interaction, converting $\bar{H}_{dd}$ into $H^{\text{sym}}_{dd}= \sum_{i<j}^N d_{ij}/3\cdot \left[ s_i^z s_j^z+ s_i^{x} s_j^{x} + s_i^y s_j^y \right] $, provided that $d_{ij}\ll \Omega$ for the Rotating Wave Approximation (RWA). For further detail refer to SM~\cite{SM}.

Rotations along $A$, $B$, $\bar{A}$, and  $\bar{B}$ can be combined in different manners, these include the frequency-switched LG sequence~\cite{LeeGoldburg1965,Mehring1972}, which corresponds to the block $A\bar{A}A\bar{A}$; and the LG4 sequence $A\bar{A}\bar{B}B$~\cite{LG4}, where if $A$ uses phase $\phi=35^{\circ}$, $B$ uses $180^\circ-\phi=145^\circ$. This is, if rotations around $A$ are exerted with $\sum_{i=1}^N \left[ \Delta s^z_i + \Omega s_i^{35^{\circ}} \right]$, for $B$, $\sum_{i=1}^N \left[ \Delta s^z_i + \Omega s_i^{145^{\circ}} \right]$. Tuning the MW control in Eq.~(\ref{eq:IP}) for either the $A\bar{A}\bar{B}B$ or $A\bar{A}A\bar{A}$ block results in the average Hamiltonian 
\begin{align}
    H_{C}= \sum_j^N [\xi'_i+\gamma_e B(t)]\, f_r\, s^C_j + \sum_{i<j} H_{dd}^{sym},
    \label{eq:effectiveH}
\end{align}
see SM~\cite{SM} for the derivation. In Eq.~\eqref{eq:effectiveH}, $f_r$ and $s_j^C$ are the reduction factor and the spin operator of the $j$-th NV projected along axis $C$, which bisects $A$ and $B$, see Fig.~\ref{fig:scheme}~b). Both quantities --$f_r$ and $s_j^C$--,  depend on the chosen configuration for the MW control block (see SM~\cite{SM} for analytical expressions). In particular, the reduction factor has a value of $f_r= 0.4292$ for the MW block $A\bar{A}\bar{B}B$, and $f_r= 0.5774$ for the block $A\bar{A}A\bar{A}$.

The sequence is completed by introducing equidistant $\pi$-pulses along a direction perpendicular to axis $C$, for example along $x$. The corresponding pulse propagator is $U_{\pi}^x = \prod_{j=1}^N (-2i)\, s^x_j$. Consequently, the full sequence reads
\begin{align}
U = \prod_{k=1}^M U^k, \quad U^k = U_{\pi}^x \: \Big[ \prod_{l=1}^m U_C(t) \Big] \: U_{\pi}^x \: \Big[ \prod_{l=1}^m U_C(t) \Big],\label{eq:U}
\end{align}
where $U_C(t)$ denotes the propagator $U_C(t) = e^{-i \int_t^{t+t_c} H_C(\tau) \, d\tau }$, with $t$ corresponding to the instant at which the given MW block starts, either $A\bar{A}\bar{B}B$ or $A\bar{A}A\bar{A}$; then, $m$ defines the number of repetitions of that MW block between $\pi$ pulses, while $M$ being the number of sequence repetitions. In the toggling frame, the $\pi$ pulses imprint a modulation function $F(t) = +1,-1$ on $s_j^C$, alternating the sign of the operator with each pulse; hence, within the Average Hamiltonian Theory approach, the effective first order Hamiltonian is
\begin{align}
\bar{H} =& \sum_j^N   \gamma_e F(t)\cdot B(t)\, f_r\, s^C_j + \sum_{i<j} H_{dd}^{sym}\nonumber \\
=& H'_{CPMG}+ \sum_{i<j} H_{dd}^{sym},
\label{eq:Heff}
\end{align}
where we define $H'_{CPMG}=  \sum_j^N   \gamma_e F(t)\cdot B(t)\, f_r\, s^C_j $ to retain parallelism with the development for the CPMG protocol presented in SM~\cite{SM}. From Eq.~\ref{eq:Heff}, it follows that the modulation $F(t)$ serves the dual purpose of coupling the NV ensemble to $B(t)$ and averaging out the disorder $\xi'_j$. The latter occurs because the protocol spans $M$ full periods of $F(t)$, during which $F(t)$ takes the values $+1$ and $-1$ for equal duration, while $\xi_j$ remain constant in time.

Then, the propagator corresponding to SHIELD protocols reads
\begin{align}
U=e^{-i \bar{H} t_s}= e^{-i H'_{CPMG} t_s } e^{-i \left( \sum_{i<j} H_{dd}^{sym} \right) t_s},
\label{eq:U2}
\end{align}
where $t_s=MT$ is the duration of the sequence. Recall the decomposition of the propagator, which is permitted due to the commutation relation $ [\sum_j^N s^C_j , \sum_{i<j} H_{dd}^{sym}]=0$.

We can now compute the spin population for $q$ interacting NVs after the application of a SHIELD protocol. For that, note that the initial $\frac{\pi}{2}$ pulse in Fig.~\ref{fig:scheme}~(a) prepares the NV ensemble in the state $\rho_0=\bigotimes_{j}^{N} (\mathbb{1}+\sigma_j^x)$, while a final pulse of an angle $\beta$ along $x$  transfers the NV population from $C^{\perp}$ to the $z$-axis for measurement, see Figure \ref{fig:scheme}.~b). Thus, following a procedure similar to that in Eq. (\ref{eq:Ft1}-\ref{eq:B13}) in SM~\cite{SM},
\begin{align}\label{eq:signal}
\langle {\textstyle \sum_{i=1}^q} s_i^z \rangle &=\text{Tr}\left[ U\rho_0 U^\dagger \sum_j^q s_j^{C^\perp}\right] \nonumber \\
&= \text{Tr}\left[ \rho_0 \left( e^{i H'_{CPMG} t_s} \sum_j^q s_j^{C^\perp}e^{-i H'_{CPMG}t_s} \right) \right] \nonumber \\
&\approx \frac{ 2\gamma_e B_t t_s}{\pi } f_r \cos(\phi_t) \  \frac{q}{2},
\end{align}
where the dipole-dipole coupling is absent, as the residual interaction does not generate any dynamics; 
i.e., since $[\rho_0, \sum_{i<j} H_{dd}^{\mathrm{sym}}] = 0$, it follows that $e^{-i \left( \sum_{i<j} H_{dd}^{\mathrm{sym}} \right) t_s} \, \rho_0 \, e^{i \left( \sum_{i<j} H_{dd}^{\mathrm{sym}} \right) t_s} = \rho_0$.

\begin{figure*}[t]
\includegraphics[width= 1 \linewidth]{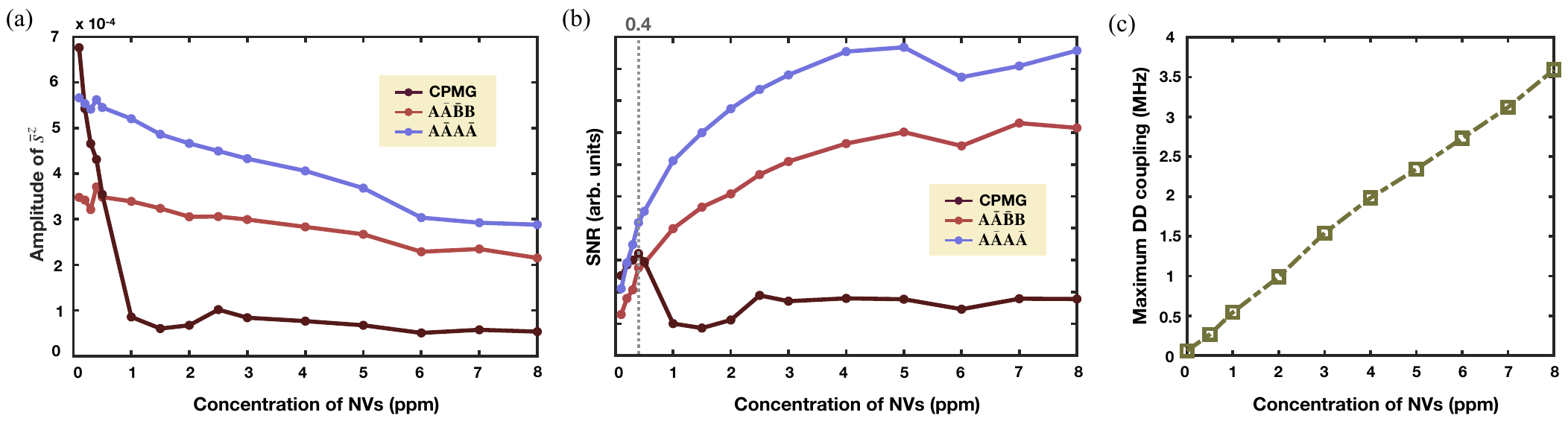}
\caption{Numerical results. In every graph, the NV density in the $x$-axis corresponds to the concentration of NV centers aligned with the external magnetic field. This direction is the preferential alignment in [111] diamond cuts~\cite{Lesik2014, Michl2014, Fukui2014, Miyazaki2014, Osterkamp2019}. (a) Oscillation amplitude of $\bar{s}^z$ for different NV concentrations. Purple curve corresponds to the CPMG protocol, while red and blue curves to SHIELD using the $A\bar{A}\bar{B}B$ and $A\bar{A}\bar{A}A$ MW blocks, respectively. (b) Expected SNR considering $\sqrt{N}$ scaling of the noise. For CPMG sequence, at $0.4$ ppm is reached the best sensitivity, while SHIELD protocols exhibit an increasing tendency of SNR for sensible concentrations. (c) Maximum dipole-dipole coupling, i.e., $d_{ij}/(2\pi)$, at different NV concentrations. This is computed by averaging the maximum dipole-dipole interaction over thousands of distinct configurations, each constituted, on average, by four NVs.}\label{fig:results}
\end{figure*}

Comparing the latter to the CPMG signal in Eq.~\eqref{eq:spectrum}, we observe two key differences. First, as a direct consequence of the continuous MW driving, our protocol introduces the reduction factor $f_r$ in the signal amplitude; second, eliminating the dipole-dipole interaction from the output signal enables all clusters to contribute constructively to the total signal. In other words, under the continuos MW driving, the term $\langle \sum_{i=1}^q s_i^x \rangle_{\rho^\ast}$ in Eq.~\eqref{eq:spectrum} converges to the upper bound $\frac{q}{2}$ for each cluster, resulting in  $\frac{N}{2}$  when accounting for the whole NV ensemble. Therefore, SHIELD, at the cost of introducing the $f_r$ factor, enables to leverage highly doped diamonds for increased sensitivity through the symmetrization of the dipole-dipole interaction, what leads to commutation relations that inhibit the detrimental expression of them.

Besides, the protocol is versatile: by adjusting the number $m$ of MW blocks between $\pi$-pulses in the sequence of Fig.~\ref{fig:scheme}~(a), enables the detection of target signals accross a broad frequency range. From a technical perspective, the delivered  $\pi$ pulses induce a modulation function $F(t)$ with a period $T=4m/\Omega$, which can be tuned changing $m$\cite{footnote1}. Then, via the first harmonic of $F(t)$, NVs couple to NMR signals of frequency $\sim1/T$, which can take values from tens of kHz to MHz. In particular, the lower range (kHz) allows for high-resolution NMR signal detection in a high-field scenario, where the sample-NV interaction is mediated by an RF field~\cite{AERIS, DRACAERIS, J_INSECT, Solid_state}; while the higher range (MHz) corresponds to medium-field NMR, where SHIELD excels integrated in heterodyne schemes~\cite{footnote2}. In contrast, other methods for highly doped diamonds~\cite{Choi2020, droid, Zhou2023, Tyler2023, Arunkumar2023, Zhou2024} are limited to a narrower frequency range in the MHz regime, limiting their application to medium field NMR.

\section{RESULTS AND DISCUSSION} To assess the performance of SHIELD, we conducted numerical simulations for the particularly interesting high-field scenario. Constrained by computational resources, simulations are limited to clusters of up to six dipolar-coupled NVs, over which distinct MW detection schemes are applied. In particular, the clusters couple to a NMR signal via $(i)$ SHIELD with a $A\bar{A}\bar{B}B$ MW block, $(ii)$ SHIELD with a $A\bar{A}A\bar{A}$ MW block, and $(iii)$ a standard CPMG sequence. To carry out the comparison between these schemes, we evaluate the average NV spin population across hundreds of clusters i.e., $Q\ge 300$) as
\begin{equation}
\bar{s}^z = \frac{1}{\sum_{k=1}^Q q_k} \sum_{k=1}^Q \sum_{i=1}^{q_k} \langle s_i^z \rangle,
\end{equation}
for systems with different NV densities. Further details are provided in SM~\cite{SM}.

The simulations are performed using Eq.~\eqref{eq:IP} as starting point, dressed with realistic noise effects. Specifically, the MW driving has an amplitude of $\Omega/(2\pi)=20$ MHz, with a $1\%$ constant Rabi error applied to both the $\pi$-pulses and MW blocks. Detuning errors in the driving are incorporated into the disorder shifts $\xi_i$, which also account for strain in the diamond. Since all NV centers within a cluster are spatially localized, we assume they share the same $\xi_i$, drawn from a Gaussian distribution with zero mean and standard deviation of $4~\text{MHz}$~\cite{Lukin}. In addition, each NV acquires an additional shift due to dipole-dipole interactions, so that the total site-dependent shift is $\xi'_i = \xi_i + \sum_{i \neq j} \frac{d_{ij}}{2}$. On the other hand, the NMR signal $B(t)$ has frequency $\omega_t /(2\pi)=95.68$ kHz and an amplitude $B_t=1$ nT. The latter corresponds to the thermal polarization of protons in a microscale pure ethanol sample under $B_z\sim 2$ T~\cite{AERIS}. Additionally, $m=32$ and $M=2$,  as the NV center decoherence time  under dynamical decoupling ($T_2\sim 20$ $\mu$s) permits the detection of two oscillations of $B(t)$, which last $t_{s}=21$ $\mu$s.

In Fig.~\ref{fig:results}, we present the numerical outcomes. The first plot (a) shows the amplitude of $\bar{s}^z$ at various NV concentrations, and the second plot (b) illustrates the corresponding SNR, accounting for shot noise and projection noise. The SNR is computed because it reflects the advantages of higher NV concentration; namely, a larger number of NV sensors significantly reduces projection noise. Details on the methodology are provided in SM.

 Analyzing the outcome for the CPMG case, in Fig.~\ref{fig:results}~(a), we observe a rapid decay in signal amplitude as NV concentration increases. This decay occurs becayse at higher density, NV centers are closer together, leading to stronger dipole-dipole interactions, see Fig.~\ref{fig:results}~(c). Consequently, in Fig.~\ref{fig:results}~(b), we observe a sensitivity sweet spot for the CPMG. This is, for low NV density, the SNR improves with NV concentration, benefiting from the increasing of NV sensors, until the dipole-dipole interaction among NVs becomes overly strong, causing a decay at 0.4 ppm. Then, according to our numerical results, 0.4 ppm is the optimal concentration for the CPMG scheme; nevertheless, further numerical analysis suggests that this would serve as an upper bound (see SM~\cite{SM}).
 
Concerning SHIELD protocols, in Fig.~\ref{fig:results}~(a), they exhibit a gradual decay in signal amplitude as NV density increases. This is because dipole-dipole interactions become stronger, making the RWA less valid. Nevertheless, this decay is slow, indicating significant suppression of dipole-dipole interactions within the considered NV densities. In this regard, note that the SHIELD sequence incorporating the $A\bar{A}A\bar{A}$  block exhibits greater amplitude compared to the $A\bar{A}\bar{B}B$ configuration, although the latter demonstrates superior robustness, characterized by a slower decay. Moving to Fig.~\ref{fig:results}~(b), this shows that the reduction factor $f_r$ makes SHIELD protocols unsuitable in scenarios where dipole-dipole interactions among NVs are negligible. However, as the NV concentration increases, the amplitude penalty from $f_r$ is compensated by the dipole-dipole cancellation provided by SHIELD. This enables to operate with a large number of NV detectors, resulting in sensitivity levels that exceed those achieved by standard CPMG at its peak performance.

To further illustrate the comparison, we focus in the case of $3$ ppm, where dipole-dipole interactions of $1.5$ MHz are observed, see Fig.~\ref{fig:results}~(c). At this concentration, CPMG suffers a drastic reduction in the signal amplitude, while SHIELD sequences are barely affected. This leads to sensitivity rates for SHIELD protocols which are five to seven times greater than that of the CPMG sequence. From a general point of view, curves in Fig.~\ref{fig:results}~(b) indicate a growing SNR for SHIELD at sensible NV concentrations, while standard schemes fail.

\section{CONCLUSION}We present SHIELD, an AC signal detection scheme tailored for highly concentrated NV-based quantum detectors. Our findings indicate that even at common NV concentrations, just above $0.4$ ppm, dipole-dipole interactions detrimentally affect standard schemes; SHIELD continuously suppresses these interactions while operating over a wide range of target signal frequencies that spans from tens of kHz to MHz. In this window, the lower frequency range is particularly interesting, as it enables the detection of high-field signals through RF mediation. Thus, SHIELD harnesses the sensitivity gains of dense ensembles and high fields, while further dipole-dipole cancellation resulting in enhanced sensitivities is possible by introducing stronger MW drivings.

\begin{acknowledgements}
A. B. U. acknowledges the financial support of the Department of Education of the Basque Government, through the initiative IKUR STRATEGY Grant No. IKUR-IKA-23/04. J. C. acknowledges the Ram\'{o}n y Cajal (Grant No. RYC2018-025197-I) research fellowship of the Ministry of Science of Spain. Authors acknowledge the Quench project that has received funding from the European Comission under The Quantum Flagship--The EU Research and Innovation Programme under Grant Agreement No. 101135742, the Spanish Government via the nanoscale NMR and complex systems Project No. PID2021-126694NB- C21, and the Basque Government Grant No. IT1470-22.

\section*{DATA AVAILABILITY}
The data generated in this study (via numerical simulations) is available from the contact author upon request.

\end{acknowledgements}

\onecolumngrid
\newpage

\clearpage
\widetext
\begin{center}
\textbf{ \large Supplemental Material}
\end{center}
\setcounter{equation}{0} \setcounter{figure}{0} \setcounter{table}{0}\setcounter{section}{0}
\setcounter{page}{1} \makeatletter \global\long\def\theequation{S\arabic{equation}}
 \renewcommand{\theHfigure}{S\arabic{figure}}
 \global\long\def\thefigure{S\arabic{figure}}
 \global\long\def\bibnumfmt#1{[S#1]}
 \global\long\def\citenumfont#1{S#1}
 
\section{The homonuclear dipole-dipole interaction among NV centers}\label{app:DD}

The general form of the dipole-dipole interaction between spins $i$ and $j$ is represented in the spin Hamiltonian by the following expression \citep{Levitt08SM}:
\begin{equation}
    H_{dd}^{ij} =b_{ij} \left[ \vec{S_i}\cdot \vec{S_j}-3(\vec{S_i}\cdot \vec{n})(\vec{S_j}\cdot \vec{n}) \right],
    \label{appeq:DD}
\end{equation}
where the interaction strength is given by the {\it dipole-dipole coupling constant}
\begin{equation}
    b_{ij}= (2\pi) \frac{\hbar \mu_0 \gamma_1 \gamma_2}{2 r_{ij}^3}.
\end{equation}
This expression is valid for systems of arbitrary spin quantum number, and in particular for NV centers, which are spin s=1 systems. In that case, $S_i$ denotes the s=1 spin operators. For NV-NV interaction, $\gamma_1=\gamma_2=\gamma_e=-28.02$ GHz/T, and the physical constants take their standard value: $\hbar=1.05457182 \times 10^{-34} \frac{m^2 kg}{s}$, $\mu_0=12.56637\times 10^{-7} \frac{kg m}{s^2A^2}$. Besides, the relative position vector between NV $i$ and $j$ is $\vec{r_{ij}}=r_{ij}\vec{n}$, then $r_{ij}$ represents the distance between them and $\vec{n}=(n_x,n_y,n_z)$ is the corresponding unit vector.

In the presence of an external magnetic field $B_z$ aligned with the geometrical axis of the NV, the free energy Hamiltonian of the NV center is $H_i^0= D (S_i^z)^2+ B_z |\gamma_e| S_i^z$, where $D= 2.99$ GHz is the zero field splitting. The eigenstates of the free energy Hamiltonian, labeled $\ket{0}_i$, $\ket{-1}_i$ and $\ket{1}_i$, have eigenvalues $\omega_{\ket{0}}=0$ and $\omega_{\ket{\pm1}}=D\pm |\gamma_e| B_z$. This is, $H^0_i= \omega_{\ket{1}} \ketbra{1}_i +\omega_{\ket{-1}} \ketbra{-1}_i$. Moving to the rotating frames of the free energy terms $H^i_0+ H^j_0$, the dipole-dipole Hamiltonian $H_{dd}^{ij}$ reads
\begin{align}
\bar{H}_{dd}^{ij}& \approx b_{ij} \Biggl\{ (1-3n_z^2)S_i^z S_j^z +e^{i (H^i_0+ H^j_0) t} \left[ S_i^x S_j^x + S_i^y S_j^y-3 (S_i^x S_j^x n_x^2 + S_i^x S_j^y n_xn_y + S_i^y S_j^x n_y n_x + S_i^y S_j^y n_y^2)\right] e^{-i (H^i_0+ H^j_0) t}\Biggl\} \nonumber \\
& \approx b_{ij} \Biggl\{ (1-3n_z^2)S_i^z S_j^z + (2-3n_x^2-3n_y^2) \frac{1}{2} [\ket{0,1}\bra{1,0}+\ket{0,-1}\bra{-1,0} + h.c.] \Biggl\} \nonumber \\
&= d_{ij} \Biggl\{  S_i^z S_j^z - \frac{1}{2} [\ket{0,1}\bra{1,0}+\ket{0,-1}\bra{-1,0} + h.c.]\Biggl\},
\label{eq:secular-dip}
\end{align}
where $d_{ij}$ is the \textit{secular dipole-dipole coupling constant:}
 \begin{equation}
     d_{ij}=b_{ij}(1-3 n^2_z)=b_{ij}(1-3\cos^2\theta).
 \end{equation}
In the first line of Eq.~\ref{eq:secular-dip}, the Rotating Wave Approximation (RWA) is applied to eliminate the mixed terms $S^zS^x$ and $S^zS^y$, since the NV centers' resonance frequency $\omega_{\ket{\pm1}}$ is much greater than the dipole-dipole coupling strength $b_{ij}$, i.e., $\omega_{\ket{\pm1}} \gg b_{ij}$. Similarly, in the second line, the RWA is applied once more, since both the sum and difference of the resonance frequencies, $\abs{\omega_{\ket{+1}} \pm \omega_{\ket{-1}}}$, are much larger than the coupling strength $b_{ij}$.
In this regard, recall that \ref{eq:secular-dip}  differs from the standard homonuclear secular dipole-dipole interaction described in \citep{Levitt08SM}. This is due to the asymmetry introduced by the strong zero field splitting, i.e., as $D \gg b_{ij}$,  the transitions $\ket{1,-1} \leftrightarrow
 \ket{0, 0}$ and $\ket{-1,1} \leftrightarrow \ket{0,0}$ are energetically suppressed.

Our work is developed considering a two dimensional subspace for each NV, namely for NV number $i$, that spanned by $\ket{0}_i,\ket{-1}_i$. This is a valid assumption because the interaction in \ref{eq:secular-dip} conserves the total spin population of states ${\ket{1}}_i$, for $i=1,...,N$. Then, once the system is initialised into an state with no population in ${\ket{1}}_i$, the dynamics remain confined in the manifold spanned by $\left\{ \ket{\alpha_1} \otimes \cdots \otimes \ket{\alpha_N} \mid \alpha_j \in \{0,-1\} \right\}$. Then, for the development of the protocol, we restrict ourselves to the suppression of the remaining dipole-dipole interaction, i.e. to the projection of  \ref{eq:secular-dip} onto the subspace, which is
\begin{equation}
 \bar{H}_{dd}^{ij} + \frac{d_{ij}}{2}(s_i^z + s_j^z) \quad  \text{where} \quad \bar{H}_{dd}^{ij}= d_{ij} \Bigl\{  s_i^z s_j^z - [s_i^x s_j^x +s_i^y s_j^y]  \Bigl\},
\label{eq:secular-dip-def}
\end{equation}
here $s_i$ are s=1/2 spin operators. We acknowledge that this is an idealized assumption, as perfect initialization is still challenging to achieve experimentally. Nevertheless, ongoing efforts are directed toward improving this aspect. Moreover, this Hamiltonian forms the basis for a number of theoretical works \cite{Lukin,droid,Zhou2023,Choi2020}, and its validity is supported by experimental evidence\cite{droid,Zhou2023,Choi2020}.

\section{The CPMG sequence in presence of dipole-dipole interactions}\label{app:CPMG}

The CPMG sequence is a widely used pulse sequence for detecting and characterising target AC signals. This protocol consists of a train of $\pi$ pulses, with a $\pi/2$ pulse to prepare the initial NV superposition state as well as to finally transfer the achieved phase into populations prior to measurement. To minimize the effect of errors on the drivings, the initial $\pi/2$ pulse ought to be $90^\circ$ shifted with respect to the subsequent $\pi$-pulses, i.e., for example, the initial $\pi/2$ pulse around $-y$, $\pi$ pulses along x and the last $\pi/2$ pulse along $x$, this is: $(\pi/2)_{-y}-[(\pi)_x]^n-(\pi/2)_x$. In such case, to estimate the evolution of the NVs' spin population, one can compute
\begin{equation}
	\langle \sum_{i=1}^N s^z_i \rangle= \text{Tr}{\left[ U \rho_0 U^{\dagger} \sum_{i=1}^N s^y_i\right]}, \label{eq:CPMG-measurement}
\end{equation}
where the propagator $U$ captures the dynamics generated whilst the application of the $\pi$-pulses and $\rho_0=\left[\bigotimes_{j}^{N} (1+2 s_j^x)\right]$ is the state of the NV ensemble prior to the application of the train of $\pi$-pulses. Note that the initial and final $\pi/2$ pulses are already implicit in this equation.

The dynamics generated during the $\pi$-pulses of the CPMG sequence can be described with Hamiltonian Eq.~(\ref{eq:IP}) by setting $\Delta=0$ while $\Omega$ is deployed stroboscopically. This is,
\begin{align}\label{eq:CPMG}
H_I/\hbar= \sum_{i=1}^N  \Omega(t) s_i^{x} + \sum_{i=1}^N \left[\gamma_e B(t) + \xi'_i \right]s^z_i + \sum_{i<j}^N \bar{H}_{dd}^{ij},
\end{align}
where the driving has been fixed on $x$. For the analytical derivation, we assume that the $\pi$-pulses of the CPMG protocol are instantaneous and ideal. In the interaction frame of the RF driving, $\bar{H}_{dd}^{ij}$ is invariant, while $s^z_i$ is multiplied by the modulation function $F(t)=\pm1$, which alternates sign with every $\pi$-pulse. Thus, the effective Hamiltonian is
\begin{equation}\label{eq:CPMG}
\bar{H_I}/\hbar= H_{\text{CPMG}}+\sum_{i<j}^N \bar{H}_{dd}^{ij}, \; \text{with}\; H_{\text{CPMG}}=\sum_{i=1}^N \gamma_e B(t) F(t) s^z_i,
\end{equation}
where the disorder terms have been neglected: as they are constant in time while the $\pi$-pulses alternate the sign of $s^z_i$, at the end of the protocol, at first order, no net effect is accumulated due to them. 

Since $[H_{\text{CPMG}},\sum_{i<j}^N \bar{H}_{dd}^{ij}]=0$, we can treat these terms separately, this is $U=U_\text{DD}U_\text{CPMG}$, where $U_\text{DD}$ and $U_\text{CPMG}$ are the propagators associated to Hamiltonian $\sum_{i<j}^N \bar{H}_{dd}^{ij}$ and $H_\text{CPMG}$, respectively. In particular, to compute the expected value for $\sum_{i=1}^N s^z_i$ (eq.\eqref{eq:CPMG-measurement} ), we can separate the dynamics generated by each to ease the calculation as
\begin{equation}\label{eq:B4}
\langle \sum_{i=1}^N s^z_i \rangle =  \text{Tr}{\left[ U_{\text{CPMG}} U_{\text{DD}}\rho_0 U^\dagger_{\text{DD}} U^\dagger_{\text{CPMG}}\sum_{i=1}^N s^y_i\right]}=  \text{Tr}{\left[ \left( U_{\text{DD}}\rho_0 U^\dagger_{\text{DD}} \right) \left(U^\dagger_{\text{CPMG}}\sum_{i=1}^N s^y_i U_{\text{CPMG}}\right)\right]}.
\end{equation}

First, we will focus on the dynamics induced by the CPMG. Let the target field $B(t)$ in $H_{\text{CPMG}}$ be a signal with frequency $\omega$, amplitude $B_{t}$ and initial face $\phi$, this is $B(t)=B_t \cos(\omega t+\phi)$. Additionally, the time separation between the mentioned $\pi$-pulses is $\frac{1}{2\nu}$, leading to a period $T=1/\nu$ on $F(t)$. Then, $F(t)$  can be Fourier expanded as
\begin{equation}\label{eq:Ft1}
    F(t) =  \sum_{\kappa\,\rm{odd}} \frac{4}{\pi \kappa} \cos\left(\nu \kappa t\right).
\end{equation}
Thus,
\begin{equation}\label{eq:Ft2}
 H_{\text{CPMG}}=\gamma_e B_t \cos(\omega t+\phi) \sum_{\kappa\,\rm{odd}} \left[ \frac{4}{\pi \kappa} \cos\left(\nu \kappa t\right)\right] \sum_{i=1}^N s^z_i.
\end{equation}
If the $\pi$-pulses are set so $\nu$ matches the frequency of the target signal, i.e. $\nu \approx \omega_t$, only the counter rotating terms of the first harmonic survive. For the rest of the exponents, $|\omega-\kappa \nu|>> \gamma_e B_t\frac{4}{\pi \kappa}$, and of course $|\omega+\kappa \nu|>> \gamma_e B_t\frac{4}{\pi \kappa}$, so they cancel out due to RWA. This is,
\begin{align} \label{eq:Ft3}
 H_{\text{CPMG}}& \approx \frac{ \gamma_e B_t}{\pi} \left[ e^{i(\omega-\nu)t}e^{i\phi}+ e^{-i(\omega-\nu)}e^{-i\phi} \right] \sum_{i=1}^N s^z_i \nonumber \\ 
 &= \frac{ 2\gamma_e B_t}{\pi} \cos(\delta t+\phi )  \sum_{i=1}^N s^z_i
\end{align}
where $\delta$ is the difference between the frequency of the target field $\omega$ and the actual frequency of the modulation function $\nu$, $\delta=\omega_t-\nu$, which we assume to be small, of the order of tens of Hertz. Assuming the protocol begins at $t=0$ and ends at $t=t_s$, its corresponding propagator is
\begin{align}
U_{\text{CPMG}}&=\exp{-i\int_{0}^{t_s} H_{\text{CPMG}}\, dt}\\
&= \exp{  -i \frac{ 2\gamma_e B_t}{\pi \delta}\left[ \sin(\delta t_s+\phi)-\sin(\phi)\right] \sum_{i=1}^N s^z_i}=\exp{ -i \frac{ 2\gamma_e B_t}{\pi \delta}\left[2\sin(\delta t_s/2)\cos(\delta t_s/2+\phi)\right]\sum_{i=1}^N s^z_i }.
\end{align}
Given that $\delta t_s<<1$, then $\sin(\delta t_s/2)\sim \delta t_s/2$; hence, the propagator can be approximated as
\begin{align}\label{eq:UCPMGend}
U_{\text{CPMG}}&= \exp{   -i \frac{ 2\gamma_e B_t t_s}{\pi }\cos(\delta t_s/2+\phi)\sum_{i=1}^N s^z_i}\approx \exp{  -i \frac{ 2\gamma_e B_t t_s}{\pi }[\cos(\phi)+ \mathcal{O}(\delta t_s)]\sum_{i=1}^N s^z_i } \\
&\approx \exp{   -i \frac{ 2\gamma_e B_t t_s}{\pi 
}\cos(\phi) \sum_{i=1}^N s^z_i }.
\end{align}
After obtaining a simplified expression for the propagator, we can compute
\begin{equation}
U^\dagger_\text{CPMG} \sum_{i=1}^N s^y_i U_\text{CPMG} = \sum_{i=1}^N \exp{ i \frac{ 4\gamma_e B_t t_s}{\pi }\cos(\phi) s^z_i}s_i^y=\cos(\frac{ 2\gamma_e B_t t_s}{\pi }\cos(\phi)) \sum_{i=1}^N s^y_i +\sin(\frac{ 2\gamma_e B_t t_s}{\pi }\cos(\phi)) \sum_{i=1}^N s^x_i. 
\end{equation}
We can now proceed with the target calculation $\langle \sum_{i=1}^N s^y_i \rangle$. Given that $\text{Tr}{\left[ \left( U_{\text{DD}}\rho_0 U^\dagger_{\text{DD}} \right)\sum_i^N s^i_y\right]}=0$, expression \eqref{eq:B4} simplifies to
\begin{align}\label{eq:B13}
\langle \sum_{i=1}^N s^z_i \rangle=\sin(\frac{ 2\gamma_e B_t t_s}{\pi }\cos(\phi)) \times \sum_i^N \text{Tr}{\left[ \left( U_{\text{DD}}\rho_0 U^\dagger_{\text{DD}} \right) s^i_x\right]}\approx \frac{ 2\gamma_e B_t t_s}{\pi }\cos(\phi) \times \sum_i^N \text{Tr}{\left[ \left( U_{\text{DD}}\rho_0 U^\dagger_{\text{DD}} \right) s^i_x\right]},
\end{align}
where, in the second equality, we apply the small angle approximation, given that $\gamma_e B_t t_s <<1$. The first term contains the chemical information, the second is an amplitude penalty introduced by the dipole-dipole interactions among the NVs. In the absence of this interaction,  i.e. $U_{DD}=\mathbb{1}$, then:
\begin{equation}\label{eq:B14}
\langle \sum_{i=1}^N s^z_i \rangle \approx \frac{ 2\gamma_e B_t t_s}{\pi }\cos(\phi) \times \sum_i^N\text{Tr}{\left[ \rho_0 s_x^i\right]} = \frac{ 2\gamma_e B_t t_s}{\pi }\cos(\phi) \times N\left(\frac{1}{2}\right).
\end{equation}
This is, the dipole-dipole interactions modulate the oscillation amplitude of the NV centers' population with the factor $|  \sum_i^N \text{Tr}{\left[ \left( U_{\text{DD}}\rho_0 U^\dagger_{\text{DD}} \right) s^i_x\right] }| \le N/2$. Then, in the best-case scenario, with the maximum factor $N/2$, dipole-dipole interactions leave the amplitude of the signal unaltered; in all other cases, these interactions lead to detrimental signal attenuation.

For example, considering two NV interaction, i.e. $N=2$, the factor introduced by the dipole-dipole interaction is
\begin{equation}\label{eq:B15}
 \sum_i^{N=2} \text{Tr}{\left[ \left( U_{\text{DD}}\rho_0 U^\dagger_{\text{DD}} \right) s^i_x\right] }= \cos(d_{12} t_s).
\end{equation}
Hence, this factor achieves a value within the range $[-1,1]$, depending on the strength of the dipolar interaction. Note that this expression is consistent with previous results, as when $d_{ij}\rightarrow 0$ (there is no dipole-dipole interaction), \eqref{eq:B15} achieves its maximum value.

\newpage
 \section{Evolution under the MW driving blocks}\label{app:LG4}
The AC detection protocol we present, incorporates MW driving blocks that continuously suppress dipole-dipole interactions among NVs. In this section, we provide the analytical derivations associated to the MW blocks to support the formulas in the main text. 
\subsection{Suppression of dipole-dipole interaction}
The MW driving blocks consist of detuned $2\pi$ rotations around distinct axis, with a specific detuning that suppresses dipole-dipole interactions. These MW rotations constitute the basis of the SHIELD protocols.

\begin{wrapfigure}{r}{0.3\textwidth}
  \begin{center}
    \includegraphics[width=0.3\textwidth]{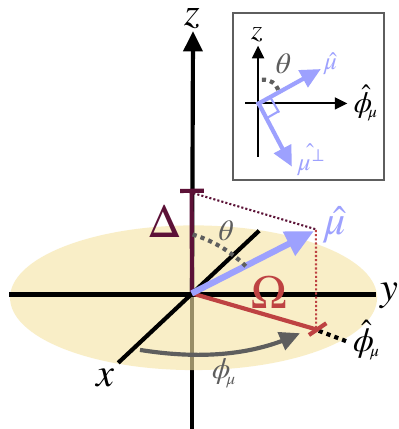}
  \end{center}
  \caption{Illustration of the effective direction of the MW driving $\hat{\mu}$, due to the Rabi amplitude in direction $\hat{\phi}$ and the detuning along $z$. The resultant angle that $\mu$ forms with $z$ is $\theta$. The upper-right inset shows the direction $\hat{\mu^{\perp}}$.}
  \label{fig:Appendix1}
\end{wrapfigure}

If we consider an NV ensemble subjected to detuned MW radiation, where the carrier frequency of the driving $\omega_c$ relates to the resonance frequency of the NV $\omega_{\ket{-1}}$ as  $\omega_c=\omega_{\ket{-1}} - \Delta$, in the rotating frame with respect to the free energy terms, the Hamiltonian is given by Eq.~\eqref{eq:IP}
\begin{align}
\frac{H_I}{\hbar}=  \sum_{i=1}^N \left[ \Delta s^z_i + \Omega s_i^{\phi} \right]+  \sum_{i<j}^N \bar{H}_{dd}^{ij} +\sum_{i=1}^N \delta_i(t) s^z_i,
\label{eq:IP3}
\end{align}
where $\delta_i(t)= \gamma_e B(t)+ \xi'_i$. In order to analyze the dynamics, it is convenient to perform a change of basis:
\begin{align}
s_i^{\mu} &= \cos(\theta)\,s_i^z + \sin(\theta)\,s_i^{\phi_\mu}, \notag \\
s_i^{\mu\perp} &= -\sin(\theta)\,s_i^z + \cos(\theta)\,s_i^{\phi_\mu}, \notag \\
s_i^3 &= 2 \left( s_i^{\mu\perp} \times s_i^{\mu} \right)
\label{eq:basis}
\end{align}
where \(\hat{\mu}\) denotes the effective direction of the MW radiation, and \(\phi_\mu\) is the angle between \(\hat{\mu}\) and the \(x\)-axis; see Fig.~\ref{fig:Appendix1}.
Thus, $\cos(\theta)=\frac{\Delta}{\sqrt{\Delta^2+\Omega^2}}$ and $\sin(\theta)=\frac{\Omega}{\sqrt{\Delta^2+\Omega^2}}$. For simplicity of expressions, we set \(\phi_{\mu} = \pi/2\). With this choice, the rotated spin components read: $s_i^{\mu} = \cos(\theta)s_i^z + \sin(\theta)s_i^y$, 
$s_i^{\mu\perp} = -\sin(\theta)s_i^z + \cos(\theta)s_i^y$ and $s_i^3 = s_i^x$. The derivation proceeds analogously for arbitrary values of \(\phi_{\mu}\).  

In the rotated basis, Eq.~\eqref{eq:IP3} can be rewritten as
\begin{align}
\frac{H_I'}{\hbar}= \sum_{i=1}^N \gamma s^\mu_i+ \sum_{i<j}^N \bar{H}_{dd}^{ij} +\sum_{i=1}^N \delta_i(t) [ \cos{(\theta)} s^\mu_i- \sin{(\theta)} s^{\mu^{\perp}} ]
\label{IP4}
\end{align}
where $\gamma=\sqrt{\Omega^2+\Delta^2}$ is the amplitude of the effective MW driving in direction $\hat{\mu}$. If one moves to a second interaction picture with respect to the driving term ($H_0^{II}=\sum_i\frac{\gamma}{2}\sigma_i^\mu$), RWA can be invoked assuming that $d_{ij}<<\gamma$ and  $|\gamma_e B(t)+ \xi'_i| <<\gamma$,   thereby eliminating the non-secular terms. In that case, the Hamiltonian reads
\begin{align}
H_{II}=&\sum_{i<j}^N d_{ij} \Biggl[(-\cos^2\theta+\sin^2{\theta})s_i^\mu s_j^\mu+ (-\sin^2\theta+\cos^2\theta+1)\frac{1}{2}[s_i^{\mu^\perp}s_j^{\mu^{\perp}}+s_i^{x}s_j^{x}] \Biggl] \nonumber \\
&+\sum_{i}^N  \delta_i(t) \cos(\theta)\, s_i^\mu.
\label{IP5} 
\end{align}

Unlike the standard spin-1/2 system scenario, when dealing with NV centers there is no angle $\theta$ that simultaneously nullifies all the trigonometric coefficients involved in the coefficients of the dipole-dipole interaction (first line of \ref{IP5}). This means that the dipole-dipole interaction cannot be completely eliminated by tuning $\theta$. However, setting
\begin{align}
     \theta=\theta_{magic}=\arccos{\frac{1}{\sqrt{3}}} \Rightarrow -\cos^2\theta+\sin^2{\theta}=\frac{1}{3},-\sin^2\theta+\cos^2\theta-1=\frac{2}{3}. 
     \label{eq:condition}
\end{align}

Eq.~\ref{IP5} becomes:
\begin{equation}
H_{II} = \sum_{i<j}^N d_{ij}\cdot \frac{1}{3} \left[ s_i^\mu s_j^\mu + s_i^{\mu^\perp} s_j^{\mu^\perp} + s_i^x s_j^x \right] + \sum_{i}^N  \delta_i(t)\cos(\theta_{magic})\, s_i^\mu,
\label{eq:remaining}
\end{equation}
where $\cos(\theta_{magic})= \frac{1}{\sqrt{3}}$. In this equation observe that the residual dipole-dipole interaction $H^{\text{sym}}_{dd}= \sum_{i<j}^N d_{ij}/3\cdot \left[ s_i^\mu s_j^\mu + s_i^{\mu^\perp} s_j^{\mu^\perp} + s_i^x s_j^x \right] = \sum_{i<j}^N d_{ij}/3\cdot \left[ s_i^z s_j^z+ s_i^{x} s_j^{x} + s_i^y s_j^y \right] $ is reduced in amplitude by a factor of \(1/3\) with respect to the original coupling, and more importantly, it takes a highly symmetric form. As a result, it commutes with a broad class of operators, in particular those of the form \(\sum_{i=1}^N s_i^\alpha\), where \(s_i^\alpha\) denotes the spin projection along an arbitrary direction \(\alpha\). Consequently, in first order approximation, this interaction does not manifest in the measured observables, as discussed in the main text.

The condition \ref{eq:condition} is translated to physical variables as follows: given that $\cos(\theta)=\frac{\Delta}{\sqrt{\Delta^2+\Omega^2}}$, when the deliberate detuning of the MW driving is $\Delta=\frac{\Omega}{\sqrt{2}}$, then $\theta=\theta_{magic}$. Hence, as long as the RWA is satisfied, i.e., $d_{ij}<<\gamma$, one should apply a deliberate detuning that relates to the amplitude of the driving as $\Delta=\frac{\Omega}{\sqrt{2}}$, which combined with DD,  avoids the effects of the dipole-dipole interaction on the measurement.

\subsection{Dynamics within the MW block}

\begin{figure*}[b]
\includegraphics[width= 0.7 \linewidth]{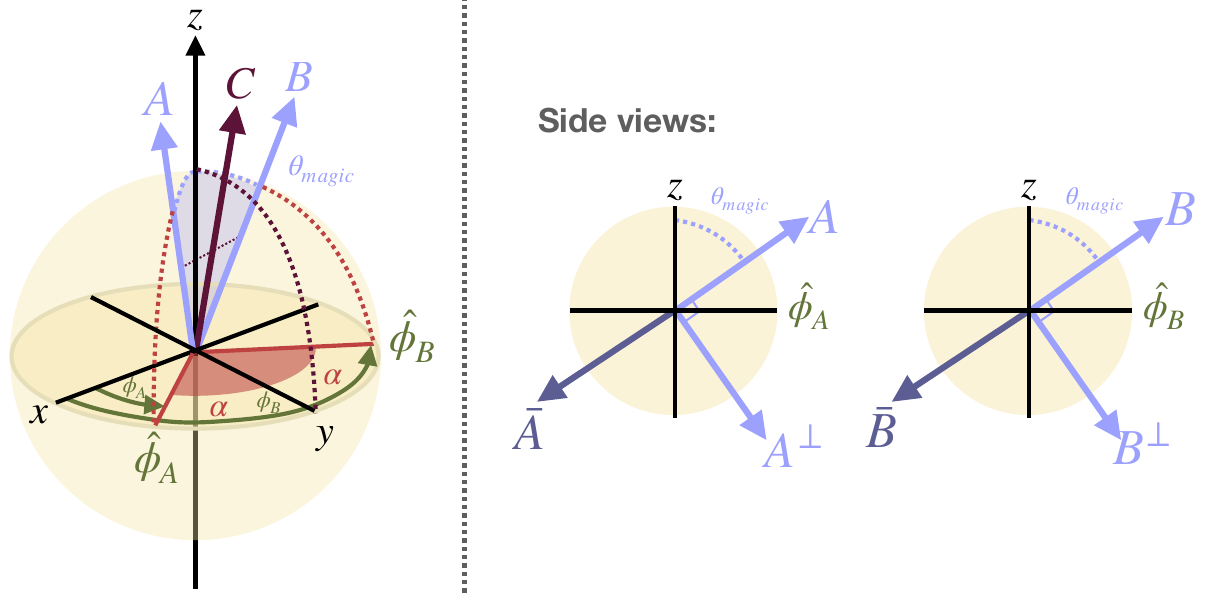}
\caption{Illustration showing the pertinent directions and angles to follow the analytical derivation. On the left, the 3D scheme, where A (B) forms with x in the XY plane an angle $\phi_A$ ($\phi_B$), defining the directions $\hat{\phi_A}$ and $\hat{\phi_B}$. On the right, side cuts of the sphere showing the planes containing $z$ and the directions $\hat{\phi_\mu}$.}\label{fig:angles}
\end{figure*}

The dipole-dipole suppressing MW block consists on four rotations: one around $A$, one around its opposite vector $\bar{A}=-A$, another around the opposite of $B$ (this is $\bar{B}=-B$) and a last one around $B$ (see Fig.~\ref{fig:scheme}). In order to see the dynamics, we should define the Hamiltonian governing during each rotation. For that, it is convenient to recycle the basis \eqref{eq:basis}, where $\mu$ can be $A$ or $B$, with the corresponding $\phi_A$ or  $\phi_B=\phi_A+2\alpha$, where $\alpha=90^\circ-\phi_A$; and $\theta=\theta_{magic}$ (see Fig.~\ref{fig:angles}).
Rewriting Eq.~\eqref{eq:IP} in terms of the operators defined in Eq.~\eqref{eq:basis}, and fixing $\theta = \theta_{\text{magic}}$, we obtain:
\begin{align}
H^{\mu}_I = \sum_{j=1}^N \gamma\, s_j^\mu + \sum_{i<j} H^{\text{sym}}_{dd} + \sum_{j=1}^N \delta_j(t) \cos(\theta_{\text{magic}})\, s_j^\mu ,
\end{align}
where $\gamma = \sqrt{\Omega^2 + \Delta^2} = \sqrt{\frac{3}{2}}\,\Omega$ is the amplitude of the effective MW driving along direction $\mu$. For simplicity, counterrotating terms proportional to $s_j^{\mu\perp}$ have been neglected under the RWA, as discussed in the previous section. On the other hand, since $s^{\bar{A}}_j=-s^{A}_j$ and $s^{\bar{B}}_j=-s^{B}_j$, the Hamiltonian governing during rotations around the opposite axes is
\begin{align}
H^{\mu}_I = \sum_{j=1}^N \gamma\, (-s_j^\mu) + \sum_{i<j} H^{\text{sym}}_{dd} + \sum_{j=1}^N \delta_j(t) \cos(\theta_{\text{magic}})\, s_j^\mu ,
\end{align}

Then,  the propagator corresponding to the MW block $A\bar{A}\bar{B}B$ is
\begin{align} 
    U_{A\bar{A}\bar{B}B}&= e^{-i \int_{t_{B}} H^B_I dt} e^{-i \int_{t_{\bar{B}}}H^{\bar{B}}_I dt} e^{-i \int_{t_{\bar{A}}}  H^{\bar{A}}_I dt} e^{-i \int_{t_A} H^A_I dt}\nonumber\\[8pt]
&=\exp\left(-i \int_{t_{\bar{B}}+t_{B}}\left[ \sum_{j=1}^N \delta_j(t)\cos(\theta_{magic}) \, s_j^B + \sum_{i<j} H_{dd}^{\text{sym}} \right] dt \right)\times \exp\left(-i \int_{t_A+t_{\bar{A}}} \left[ \sum_{j=1}^N \delta_j(t)\cos(\theta_{magic})\, s_j^A + \sum_{i<j} H_{dd}^{\text{sym}} \right] dt \right)
\label{eq:AABB}
\end{align}
where $t_{\mu}$ denotes the time slot of duration $1/\gamma$ during which a $2\pi$ rotation around axis $\mu$ is applied.  
If we assume that the magnetic field does not vary significantly during consecutive MW rotations, we can approximate
\[
\int_{t_A + t_{\bar{A}}} \delta_j(t)\, dt \quad \text{and} \quad \int_{t_B + t_{\bar{B}}} \delta_j(t)\, dt
\]
by
\[
\frac{1}{2} \int_{t_A + t_{\bar{A}} + t_B + t_{\bar{B}}} \delta_j(t)\, dt.
\]
Moreover, since $|\delta_j(t)| \cdot \frac{1}{\gamma} \ll 1$, the Baker--Campbell--Hausdorff formula can be truncated at first order.  
This allows to approximate Eq.~\ref{eq:AABB} as follows
\begin{align}
    U_{A\bar{A}\bar{B}B}&\sim \exp{-i \left[ \sum_j^N \left(\int_{t_A+t_{\bar{A}}+t_B+t_{\bar{B}}} \delta_j(t) dt \right) \cos{(\theta_{\text{magic}})} \frac{(s^A_j+ s^B_j)}{2}+ \int_{t_A+t_{\bar{A}}+t_B+t_{\bar{B}}}  \sum_{i<j} H_{dd}^{sym} dt \right]}.
\end{align}
Given that $(s^A_j+ s^B_j)/2= \cos{(\theta_{\text{magic}})}s^z_j + \sin{(\theta_{\text{magic}})}\sin\phi_A s^y_j\equiv s^C_j \sqrt{\cos^2{(\theta_{\text{magic}})}+\sin^2{(\theta_{\text{magic}})}\sin^2{\phi_A}}$, we can define $C$ as the direction of the axis along which the effective MW rotation occurs when radiating the MW block fast enough (i.e. when the aforementioned conditions are met). In that case, the effective Hamiltonian for the MW block is
\begin{align}
    H_{C}= \sum_j^N [\xi'_i+\gamma_e B(t)]\, f_r\, s^C_j + \sum_{i<j} H_{dd}^{sym}
\end{align}
where $f_r=\sqrt{\cos{(\theta_{\text{magic}})}^2+\sin{(\theta_{\text{magic}})}^2\sin\phi_A^2} \cos{(\theta_{\text{magic}})}$. This reduction factor is a consequence of the applied continuous driving, which limits the NV centers from fully detecting the target signal, thereby introducing an amplitude penalty. In other words, it is the price to pay in exchange of protection against dipole-dipole interactions, and it depends on the specific sequence. In particular, for the MW block  $A\bar{A}\bar{B}B$, $\phi_A=35^{\circ}$, then $f_r=0.4292$; while for the frequency switched protocol $A\bar{A}A\bar{A}$, $\phi_A=90^\circ$, resulting in a reduction of $f_r=\cos{\theta_\text{magic}}=1/\sqrt{3}$.

\section{Numerical simulations}\label{app:numerics}
In this work, via numerical simulations, we compare the performance of three AC detection protocols, these are: CPMG, SHIELD utilizing MW block $A\bar{A}A\bar{A}$ and SHIELD with $A\bar{A}\bar{B}B$. Fig.~\ref{fig:results} in the main text, presents the final numerical outcomes; and in this section we show the methodology employed to extract the data depicted in that figure.

\subsection{High-field detection sequence}

\begin{figure*}[h]
\includegraphics[width=0.8  \linewidth]{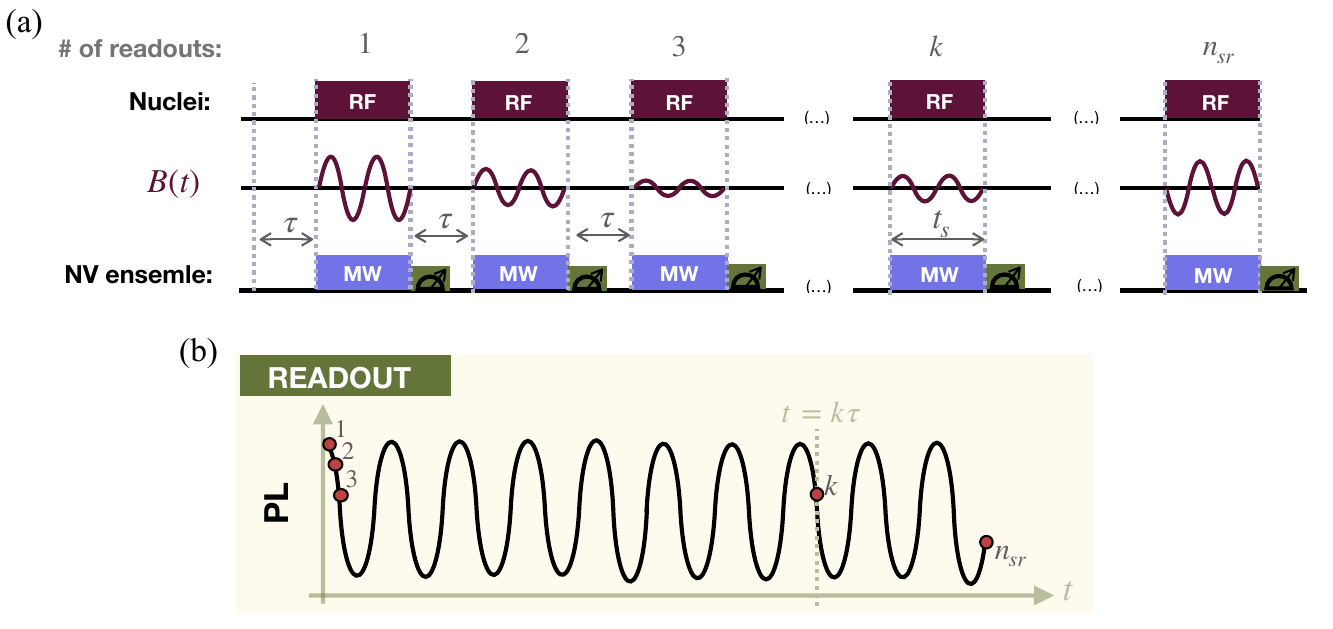}
\caption{(a) Illustration of the AERIS detection scheme, which consists of $n_{sr}$ measurements. In each measurement, a 2$\pi$ rotation is applied to nuclei in the NMR sample via RF, inducing the field $B(t)$ on the NV ensemble. This field then is simultaneously detected by the NV ensemble using a MW measuring protocol that can be any of the three mentioned in the text. (b) Schematic of the recorded PL signal with the detection scheme. Each red point corresponds to an NV ensemble readout, resulting in oscillations with frequency $\delta$. }\label{fig:app2}
\end{figure*}

To compare the performance of these measurement protocols, we integrate them within the detection scheme AERIS~\citep{AERISSM}, as they would be employed in high field experimental setups. AERIS is a scheme tailored for high field signals that consists on multiple short-duration NV measurements. In this manner, the NV decoherence does not limit the spectral resolution; instead, the frequency resolution is dictated by the decoherence of the target signal, just as in heterodyne schemes. During the measurements within AERIS, the NV ensemble captures the RF-induced Rabi oscillations, which encode chemical information of the NMR sample in the amplitude. In particular, the NV ensemble detects the longitudinal nuclear Rabi signal, and for that [111] cut diamonds are required. In the simulations, the Rabi signal is detected with one of the three protocols --CPMG, SHIELD utilizing MW block $A\bar{A}A\bar{A}$ or SHIELD with $A\bar{A}\bar{B}B$--, and to ensure a fair comparison, the measurement duration is kept identical across all cases. Refer to Fig.~\ref{fig:app2} for the overall detection scheme.

For the sake of simplicity, we consider a single chemical shift $\delta=200$ Hz as the target parameter. Then, one can show that the magnetic field on the NV ensemble during the k-th magnetometry measurement reads as
\begin{equation}
B^k(t)= B_t^k \sin(\omega_t t) \qquad	\text{ where } \qquad B_t^k=B_t \cos(\delta\tau k),
\end{equation}
where $\tau$ represents the time interval between successive measurements. This is, the amplitude of the magnetic field sensed by the NV ensemble changes from one measurement to the next one; in particular, it varies sinusoidally according to the chemical shift $\delta$. Then, the expected response of the NV centers, using Eq.~\eqref{eq:spectrum} and \eqref{eq:signal} of the main text, is the following
\begin{equation}\label{eq:AERIS}
\langle \sum_{j=1}^q s_z\rangle^k= \frac{2\gamma_e B_t \cos( \delta\tau (k-1) ) t_s f_r}{\pi} \times \langle \sum_{j=1}^q s_j^x \rangle_{\rho^*},
\end{equation}
where q is the number of NV centers in the cluster. When the selected measuring protocol is the CPMG, the term $\langle \sum_{j=1}^q s_j^x \rangle_{\rho^*}$ has a varying value in between $-q/2$ and $q/2$ depending on the dipole-dipole interactions within the cluster and $f_r=1$. For the SHIELD protocol, $\langle \sum_{j=1}^q s_j^x \rangle_{\rho^*}=q/2$, and $f_r=0.4292$ for the $A\bar{A}\bar{B}B$ block and $f_r=0.5774$ for the $A\bar{A}A\bar{A}$. 

Then, performing multiple measurements that take place at different time instants $\tau k$, leads to the reconstruction of the sinusoidal behaviour $\cos( \delta\tau k )$ in the PL signal, see Fig.~\ref{fig:app2}~(b), from where the parameter $\delta$ can be estimated. The amplitude of the PL signal, and thereby the sensitivity, varies depending on the particular protocol employed for the NV measurement combined with environmental conditions, such as the NV concentration and the strength of the decoupling drivings.

\subsection{Methodology}
In order to compare the performance of the detection protocols, we require a representative signal for each density and protocol. This is challenging given the computational constraints: the simulations are restricted to clusters of a few interconnected NV centers. Then, to generate reliable representative signals, we average the output NV signal across hundreds of random clusters, continuing until the average signal remains stable, indicating that convergence is reached. This approach, within our computational limitations, captures the randomness of NV center positioning and consequent diversity of dipole-dipole coupling structures that give rise to heterogeneous signals.

To create each random cluster, we place 30 NV centers following uniform probability distribution within a volume that would correspond to 30 NV centers if these were periodically arranged. This volume is 
\begin{equation}
    V (\text{ppm})= (\frac{30\times 10^6}{8\times\text{ppm}})^{1/3}\times a^3,
\end{equation}
where $a=3.57\times10^{-10}$m is the lattice constant of diamond. Besides we impose the restriction that two NV centers cannot occupy the same lattice cell. We then select those NV centers that have fallen within a spherical volume that contains, on average, $4$ NVs. The radius of this sphere is given by
\begin{equation}
    r(\text{ppm})=( \frac{4\times10^6}{8 \times \text{ppm}} \times \frac{3}{4\pi} ) ^{1/3}.
\end{equation}
Only the configurations containing one to six NV centers are taken as valid; with this handful of NVs randomly distributed, simulations are conducted using Eq.~\ref{eq:IP} as starting point and with the parameters specified in section III of the main text. Besides $\delta=200$~Hz, $\tau=10^{-4}$s and $n_{sr}=100$. To obtain the representative signal of an NV embedded at a given density, the signals from NVs in different clusters is averaged as $\bar{s}^z= \left(\sum_{k=1}^Q \sum_{i=1}^{q_k} \langle s_i^z \rangle \right) / \sum_{k=1}^Q q_k$, where $Q$ represents the number of clusters and $q_k$ the number of NVs in the $k$-th cluster. The number of clusters included in this averaging is progressively increased until convergence, defined as reached when the mean signal varies by $<5 \%$ upon adding the last 50 configurations.

\begin{figure*}[h]
\includegraphics[width= 0.9 \linewidth]{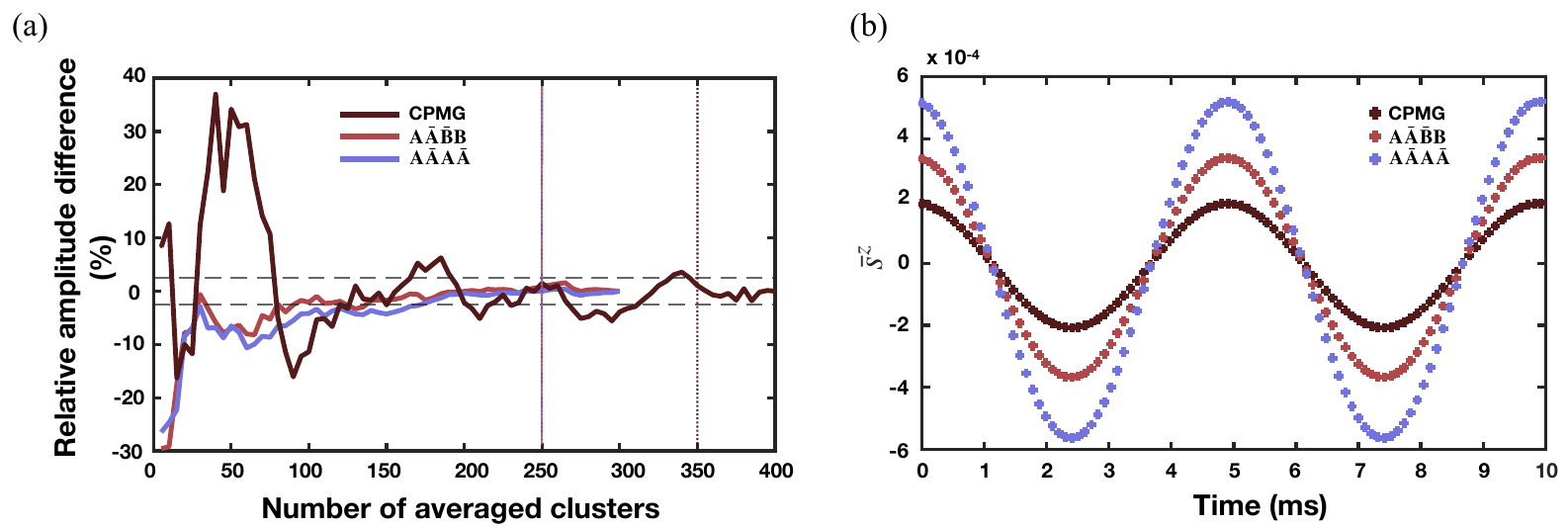}
\caption{Simulations of the model at 1 ppm NV concentration. (a) Convergence graph: the $y$-axis shows the difference between the signal amplitude obtained with a given number of averaged clusters (shown on the $x$-axis) and the converged signal amplitude, expressed as a percentage of the latter. The vertical dashed lines, using the same color code as the curves, mark the point from which the last 50 clusters are added to the averaging for each protocol. From that point onward, the amplitude difference remains within $\pm 2.5 \%$ (horizontal dashed lines) for the three cases, indicating convergence. (b) Converged signals representing the signal captured via AERIS using CPMG or SHIELD with MW block $A\bar{A}\bar{B}B$ or $A\bar{A}A\bar{A}$. For CPMG $400$ clusters are averaged, while for the other two, $Q=300$.}\label{fig:1ppm}
\end{figure*}

For illustration, Fig.~\ref{fig:1ppm} shows the convergence graph and the resulting converged signal for an NV density of 1~ppm. In the convergence graph Fig.~\ref{fig:1ppm} (a), CPMG takes longer to converge, as the averaged amplitude exhibits stronger fluctuations. This arises from the diverse dipole-dipole structures across clusters, which produce signals with varying amplitudes. In contrast, SHIELD protocols suppress these interactions, reducing heterogeneity and allowing faster convergence. As shown in Fig.~\ref{fig:1ppm}(b), all detection schemes display the expected oscillatory behavior of Eq.\ref{eq:AERIS}, with their frequencies directly encoding the target parameter $\delta$. The outputs differ only in amplitude: for CPMG, the heterogeneity among cluster signals leads to destructive interferences resulting in amplitude reduction.

In order to construct Fig.~\ref{fig:results}~(a), which portrays the amplitudes of the converged signals, convergence was achieved using the three protocols in ensembles containing one to six NV centers, for NV densities of $0.1$, $0.2$, $0.3$, $0.4$, $0.5$, $1$, $1.5$, $2$, $2.5$, $3$, $3.5$, $4$, $5$, $6$, $7$, and $8$ ppm (these concentrations correspond to NV centers aligned with $B_z$). To reach convergence, the scheme using CPMG was run over $Q=300$ clusters for densities 0.1 ppm and 0.3 ppm, over $Q=400$ clusters for 0.5 ppm, 1 ppm and 1.5 ppm and over $Q=1400$ clusters for densities above 1.5 ppm. On the other hand, the scheme using SHIELD protocols is run over $Q=300$ for every density. To obtain Fig.~\ref{fig:results}~(b),  the SNR is estimated by multiplying the values on Fig.~\ref{fig:results}~(a), which represent the response associated with the spin projection of a single referential NV center embedded in an ensemble, by $\sqrt{N}$, where $N$ is the number of NV centers in the active volume $V$ of the diamond. This is taken as $V_{\text{act}}=3.14\times 10^{-15}m^3$ \cite{GlennBucher2018SM}, thus $N (\text{ppm})= \text{ppm} \ \frac{8 \times 10^{-6}}{a^3} \ V_{\text{act}}$. This scaling, $\sqrt{N}$, corresponds to spin-projection-noise and shot-noise (i.e. the SNR only accounts for this type of noise), and illustrates the benefit of having more NV centers.

\subsection{Simulations with larger clusters}
The reduced number of constituents we can integrate into the numerical simulations alters the results. For example, the oscillatory behaviour exhibited in Fig.~\ref{fig:results}~(a) for the sequence with the CPMG protocol, we argue it is a consequence of the restricted amount of interacting NV centers in the simulations; for an infinite amount of interacting NVs we expect a smooth decay. Additionally, we hypothesize that this numerical condition results non-favorable to our protocol when performing the comparison. This is: the observed advantage of the dipole-dipole protected protocols over the CPMG-integrated version is likely a lower bound. When a bigger volume is considered and more elements are included in the clusters, the potential variations in the arrangement of NVs increase. This leads to a greater diversity of signals across different clusters, resulting in a more pronounced destructive interference in the overall signal. From a more physical point of view, one can think that more interaction channels lead to a faster decay. This phenomena affects much less in the schemes with dipole-dipole suppression, suggesting that the comparison carried out is indeed a lower bound of the improvement in sensitivity offered by SHIELD.

\begin{figure*}[h]
\includegraphics[width= 0.7 \linewidth]{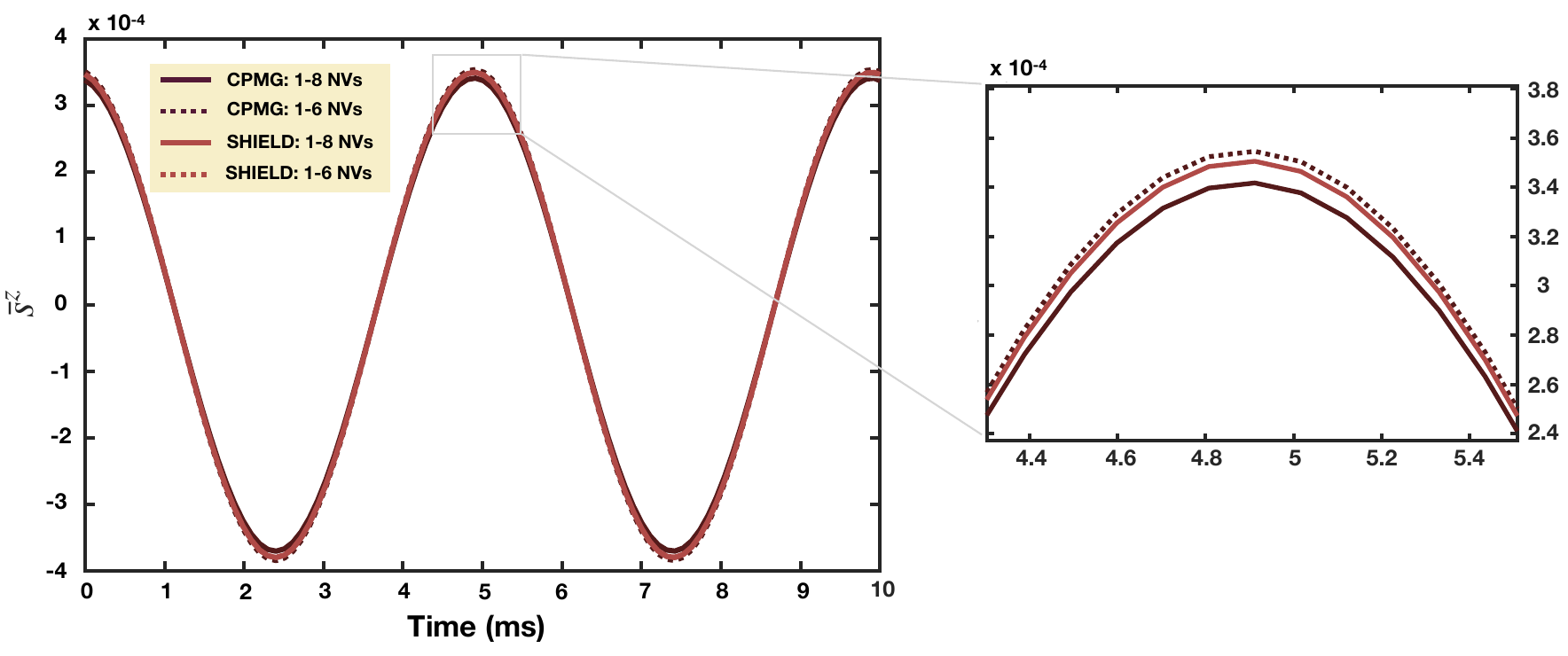}
\caption{Converged signals at $0.5$ ppm NV concentration. Continuous lines correspond to clusters comprising one to eight NV centers, while dashed lines indicate the restricted case of one to six NV centers. The color code is consistent with previous figures: purple lines denote the detection scheme that uses CPMG and red lines correspond to SHIELD with the MW block $A\bar{A}\bar{B}B$. The red dashed line coincides with the continuous one and is therefore not visible.}\label{fig:1ppm-8NVs}
\end{figure*}

In order to gain insight regarding the latter assumption, we run for $0.5$ ppm of NV concentration extra simulations with the exact same conditions as before, but with slightly greater clusters. In particular, to construct the clusters, we select the NV centers that fall in a spherical volume that in average captures six NV centers instead of four, this is
\begin{equation}
    r'\left(\text{ppm}\right)=\left( \frac{6\times10^6}{8 \times \text{ppm}} \times \frac{3}{4\pi} \right) ^{1/3},
\end{equation}
and we take as valid those configurations containing one to eight NV centers. Then, we average the NV signal over these configurations. Convergence was reached after 500 clusters, with the criterion that the averaged signal varied by less than 1\% over the last 50 realizations. The converged signals are shown in Fig.~\ref{fig:1ppm-8NVs}, where, for comparison, are also included those corresponding to the smaller clusters. The results reported in this figure support our hypothesis: the outcome signal for the SHIELD protocol remains unaffected when more NV centers are considered in the simulations. In contrast, for the protocol using CPMG, we observe that expanding the system yields smaller signal amplitude, even when the concentration remains unchanged. In particular, we observe a decay of approximately 4 \%, which causes the CPMG signal, initially slightly higher than that of SHIELD, to drop slightly below it.

\subsection{Phase noise in the MW driving}
Random fluctuations in the phase of the MW driving can degrade the NV-based measurement \cite{Berzins24SM} To assess the impact of such noise in our protocol, we conducted an analysis through additional simulations.

In the frequency domain display of the MW driving signal, phase noise manifests as broadening of the spectral peak at the carrier frequency. Motivated by \cite{Berzins24SM}, we employed as reference the performance of the commercial MW generator SRS SG386. According to its user manual, it generates a 1 GHz MW signal with the following spectral density: at offsets $10$ Hz, $1$ kHz, $20$ kHz and $1$ MHz, the power of the noise is $-80$ dBc/Hz, $-102$ dBc/Hz, $-114$ dBc/Hz and $-124$ dBc/Hz, respectively. We then extrapolated the spectral density to a carrier frequency of $60$ GHz (the approximate resonance frequency of NVs at 2 T) by applying a 6 dB/octave scaling as specified in the manual.
Next, integrating the spectral density from $10$ Hz to $1$ MHz to compute the phase jitter, resulted in an RMS jitter of approximately $160$ mrad per clock period. Finally, incorporating this phase error into the simulation of SHIELD with the MW block $A\bar{A}\bar{B}B$ (with $M=2$ and $m=32$) yields propagators with fidelities above 0.98 with respect to the ideal propagator. Thus, we conclude that MW phase noise has negligible effects on our protocol.


\begin{thebibliography}{X}
\bibitem{footnote1}
The interpulse timing, and consequently $T$, can also be adjusted by tuning the duration of the MW block $t_c$, which is determined by the driving amplitude $\Omega$ and the specific MW block design. However, it is important to note that tuning $t_c$ --for instance, by adjusting $\Omega$--, also impacts the protocol's decoupling capability.
\end{thebibliography}
\end{document}